# An Analytical Theory for the Early Stage of the Development of Hurricanes: Part I

By

Chanh Q. Kieu

Department of Meteorology, University of Maryland

College Park, Maryland

July 2004

# Table of Contents

# Abstract

In this work, a theoretical formulation for the early stage of hurricane development, in which analytical solutions have been found for this stage, is proposed. These solutions are not only consistent with observations but also offer some new insights into hurricane properties. Traditionally, theoretical approaches of hurricanes were based on incipient balanced relationships between wind and mass fields or on some scaling values, e.g. radius of maximum wind, tangential wind…, to linearize the primitive equations. Even in this linearized form, it has been not possible to come up with any analytical solutions. This is the first time (to my knowledge) an analytical theory for the initial growing of hurricanes is proposed for which analytical solutions have been found, based on an assumption of a positive feedback process of a self-induced developing system. Starting with a very simple linear theory and no friction, this theoretical theory is then advanced to a higher level in which nonlinear terms are included. The effect of friction will be considered in part II of this series of papers. Analytical results are consistent in many aspects with observations. This work will answer the question of "*how hurricanes will evolve with time*" in the presence of favorable conditions for hurricanes to grow.

# Acknowledgment

I would like to thank Dr. Dalin Zhang for his many helpful corrections. I am also grateful to Mr. Wallace Hogsett and Emily Becker for their constructive comments. This research was supported by Vietnam Education Foundation (VEF). Had it not been for their support, I would not have been able to concentrate my mind on this work.

# 1. Introduction

A traditional approach to hurricane dynamics, generally, can be divided into two categories. The first is a theoretical approach. The most preferred theoretical idea is the one in which some scaling parameters (such as radius of maximum wind speed, tangential wind speed,…) are used to form some non-dimensional numbers or to make scaling evaluations (Willoughby 1979; Montgomery and Farrell 1993; Noland and Montgomery 2002; Shapiro *et. al.* 1993; Holland 1997; etc.). Since these nondimensional numbers are small, the field variables are expanded as a series in terms of these numbers. By restricting the expansion to the first several approximations, the primitive equations are linearized. It is therefore possible to separate the mean balanced fields from the total fields and a new system of linearized equations for perturbation is considered. One of the basic characteristics of this theoretical approach is that it is assumed **a priori** that a mature hurricane has been formed so that the scaling numbers are available. Even in this case, we still don't have any information about the analytical solutions of the linearized system because of the high order and mixed derivatives between variables. This linearization was proven to be a good way to describe some hurricane characteristics at the mature stage and continued to be the preferred method until now. Another interesting theoretical idea was based on some initial equilibrium relationships between mass and wind fields, which was investigated originally by Sawyer (1947) and Yanai (1961), (see Palmen and Newton 1969, chapter 15). If this equilibrium is disturbed, tangential and radial wind will develop. By following this route, Yanai has obtained a final equation which contains several stability coefficients. Because these coefficients have no explicit expression, no exact solution was obtained thus far.

The second category is modelling. In recent decades, because of the rapid improvement of computer speed, this promising approach, in which the full system of the primitive equations has been used to simulate hurricane development, has emerged and is used extensively. This method offers a good chance (and perhaps the best way) to understand the development, structure and movement of hurricanes and continue to become more and more sophisticated (Zhang *et. al.* 2000; Challa *et. al.* 1998; Nguyen *et. al.* 2001; Zhu *et. al.* 2001; to name only a few). These models can capture very detailed processes of a hurricane, which are reasonably consistent with observations. It is however common to utilize a bogus vortex in hurricane modelling and this vortex is then integrated with time (Liu *et. al.* 1998; Pu *et. al.* 2001; Kwon *et. al.* 1999; etc.). By using numerical models, we are automatically unable to get any information about analytical solutions of hurricane processes.

So far, neither of these above approaches, theoretical approaches or modelling, gives us an exact solution of the primitive equations at the early development stage of hurricanes. Due to the nonlinear nature of system of the primitive equations as well as unknown small-scale processes, it appears that there exists no analytical solution for hurricane phenomenon to date. If we look back to many meteorological phenomena, such as fronts, stability, waves… or even Elnino, it will be found quickly that there is at least one analytical model that gives us some analytical solutions to investigate. The purpose of this work is to fill out this gap in hurricane research by setting out a framework for investigating analytical solutions at the early stage of hurricanes. This work is fresh in two ways: first it describes ***the behaviours of hurricanes at the initial time*** and secondly, it gives us an ***analytical solution for this stage***. It will be immediately anticipated that this preliminary work will contain in it many simplifications and even somewhat unrealistic assumptions. However, as we shall see, it is very useful at capturing some basic characteristics of hurricanes with radius and height, such as wind and geopotential fields. This work should be regarded as an intermediate step to a further complete theory of hurricanes. It is necessary to have such a complete hurricane theory instead of borrowing a picture from the Rankine 2D vortex (Smith 1980; Kuo *et. al.* 1999; Holland). This is not very helpful because the Rankine vortex has a real existence of an interface (i.e. water surface) while in the atmosphere it is impossible to see such a surface directly. We will find in section 3.1 that such vortex will also lead to an unrealistic description of hurricanes.

## 2. Starting equations and basics assumptions

At the most basic level, all meteorological phenomena are described by six basic equations: three momentum equations, the continuity equation, the thermodynamical equation and the equation of state. For each specific meteorological phenomenon such as waves, stabilities, fronts, etc., there are from observations some approximations which are particularly applicable to this phenomenon. Using these approximations it is then possible to obtain a system of simplified equations to investigate in detail. For hurricanes, we do the same. However, unlike the traditional approaches to hurricanes as mentioned in the introduction section, we now will attempt to obtain exact solutions of hurricane development rather than assuming the existence of a hurricane as these former approaches did. For completeness, the six basic equations of fluid motion in the atmosphere are re-written:

$$\frac{du}{dt} = -\frac{1}{\rho}\frac{\partial p}{\partial x} + F_{cx} + F_x \tag{1}$$

$$\frac{dv}{dt} = -\frac{1}{\rho}\frac{\partial p}{\partial y} + F_{cy} + F_y \tag{2}$$

$$\frac{dw}{dt} = -\frac{1}{\rho}\frac{\partial p}{\partial z} + F_{cz} + F_z \tag{3}$$

$$\frac{d\rho}{dt} + \rho.div\vec{V} = 0 \tag{4}$$

$$C_p \frac{dT}{dt} - \frac{RT}{p}\frac{dp}{dt} = J \tag{5}$$

$$p = \rho RT \tag{6}$$

where:

$$\frac{du}{dt} = \frac{\partial}{\partial t} + u\frac{\partial}{\partial x} + v\frac{\partial}{\partial y} + w\frac{\partial}{\partial z}$$

All variables in the above equations have the usual meteorological meaning. It should be noted that eddy forcings are not contained in this system because all of these variables will not be averaged as usually encountered in meteorological problems. Only after we employ averaging operators do the eddy terms appear. The reason why we need to average all variables in numerical models is simply because we don't and, more correctly, have no way to deal with instantaneous values of field variables such as P, u, v, T. This is due to the limitations of observation and the stochastic nature of the atmosphere. It is also unnecessary to require such instant values in daily life. For the theoretical approach, this average is neglected. $F_x$ and $F_y$ here are the frictional forcings only, not related in any way to the eddy terms. $F_{cx}$, $F_{cy}$, $F_{cz}$ are the Coriolis forces in the x, y and z directions, respectively. Our purpose is, starting from this system and some basic assumptions, to find out analytical solutions of wind, pressure, temperature fields which are consistent with the observations of hurricane characteristics.

Now, we need to make some basic assumptions to start with. The first assumption will be based on one important characteristic of developing systems: a positive feedback. This positive feedback at the early stage is a requirement for any unstable system to grow; otherwise, the system will come back to its initial state and nothing can occur. If we look back to our starting system and assume at the beginning that there is no motion so that the pressure surface is flat, the wind fields are zero (u=v=w=0), then a question is "how to trigger this system to evolve with time?" If one writes a program for this system and the initial data is a quiet atmosphere like that, clearly it is impossible to have a hurricane since the right hand side of the primitive system ((1) to (6)) is zero everywhere. Now I make two basic assumptions here: 1) *heating source (J) will be proportional to the vertical motion (w),* *given by: J = kw* and 2) *there is an initial vertical motion in some region.* In nature, it is obvious that although there are upward motions always and nearly everywhere but hurricanes are rarely formed. So, the relationship J=kw is not always applicable. Even if this relation is valid, there is no guarantee that a hurricane will appear. A typical example is a deep convective region with a tower of cumulus or nimbus clouds. Whether a hurricane will appear depends on many other factors as pointed out in previous works and will not be mentioned here. But whenever there is a hurricane, we can expect that this relationship exists. This relationship can be verified by comparing the vertical profile of heating from observations, e.g. in Rodgers *et. al.* (1998), and of vertical motion, e.g. in Palmen and Newton (1969). Of course, the development of a hurricane cannot proceed endlessly because when it reaches some limitation, an unknown mechanism will appear to keep the hurricane from developing further. Here, we are considering the early development of hurricanes and the assumption of positive feedback will be acceptable *provided that the final results are applied only for some limited initial period.* With the purpose of investigating the early phase of hurricane development, this assumption is made and our task now is to prove that analytical solutions describing the wind or geopotential fields of a hurricane can be found, based on this assumption. The meaning of this relation is clear. More upward motion will lead to more latent heat release and buoyancy force will be larger so that upward motion is intensified. Several common assumptions will also be used, such as Boussinesq and f-plane approximations. It will be stated explicitly whenever these approximations are used. In Part I, we will consider two theories: linear and non-linear. These theories don't have friction included. In Part II of this series of paper, friction will be added.

A good analytical theory must have solutions which satisfy at least the following properties obtained from observations:

1A) The U-shape of geopotential at all levels in the core region

1B) Convergence of radial wind below (vertical wind increases with height) and divergence of radial wind above (vertical wind decreases with height)

1C) Motion is cyclonic and geopotential is of the U-shape in the upper atmosphere, where divergence occurs (this property is very important)

1D) The profile of tangential wind needs to reach a maximum at some radius and then decrease with radius

1E) Tangential wind has to decrease with height above the boundary layer.

These observations will be used to verify whether a theory for hurricane is good enough.

## 3. Approach to analytical solutions

This section consists of two hypothetical theories. Starting from the simplest idealization in which neither nonlinear terms nor friction are considered for the linear theory in section 3.1, nonlinear terms are then added for the nonlinear theory in section 3.2.

**3.1. Linear theory**

It is quickly seen that we are here confronted with a very difficult problem: a nonlinear system ((1) to (6)). Moreover, several field variables appear simultaneously and usually it is not possible to solve this system completely analytically. It is very common to investigate a complicated system by starting first with a simplified problem. The simple description is always important for us, allowing for some preliminary understandings of behaviors of solutions. This simple problem can show us the advantages and the shortcomings of our simplifications, from which we can make a further improvement. It is first to start with the case in which all the nonlinear terms in the system ((1) to (6)) are set to zero. As we shall see, such simplified system will be helpful when the nonlinear cases are considered. The assumptions for this simplest linear theory will be:

- No non-linear terms such as $u\partial u/\partial x$, $u\partial v/\partial x$, etc. This assumption is reasonable because at the early stage of hurricane development, it may be expected that u is small such that $u\partial u/\partial x$ will be an order of magnitude smaller than $\partial u/\partial t$. In section 3.2, these nonlinear terms will be included.

- No friction. This assumption seems to be inconsistent because, in a real hurricane, the frictional convergence at low level actually seems to be the source of latent heat and upward motion. Without friction, it seems that hurricanes are not able to develop. However, as we shall see, the assumption of the proportional relation between heating and vertical motion (J=kw) makes this no-friction assumption become acceptable. More complete treatment of the role of friction will be given in Part II.

- Incompressible fluid dρ/dt=0 (the changing of density with height does not have a significant influence and will be neglected)
- Boussisnesq approximation
- Heating source (J) is proportional to vertical motion (w) (J= kw where k is a constant proportional coefficient). This assumption actually is a complementary for the no-friction assumption above. The assumption k = constant will be relaxed in an upcoming paper.
- F-plane

With these above assumptions, the primitive system now becomes:

$$\frac{\partial u}{\partial t} = -\frac{1}{\rho_0}\frac{\partial p}{\partial x} + fv \quad (1.1)$$

$$\frac{\partial v}{\partial t} = -\frac{1}{\rho_0}\frac{\partial p}{\partial y} - fu \quad (1.2)$$

$$\frac{\partial w}{\partial t} = -\frac{1}{\rho}\frac{\partial p}{\partial y} - g \quad (1.3)$$

$$\frac{\partial u}{\partial x} + \frac{\partial v}{\partial y} + \frac{\partial w}{\partial z} = 0 \quad (1.4)$$

$$C_p(\frac{\partial T}{\partial t} + w\frac{\partial T}{\partial z}) + wg = J \quad (1.5)$$

Instead of dealing with temperature, (1.3) will be re-written in terms of buoyancy: b=(T-$T_e$)/$T_e$ where $T_e$ is the temperature of environment. Introducing some new parameters, the above equations become:

$$\frac{\partial u}{\partial t} = -\frac{\partial \Phi}{\partial x} + fv \tag{1.1'}$$

$$\frac{\partial v}{\partial t} = -\frac{\partial \Phi}{\partial y} - fu \tag{1.2'}$$

$$\frac{\partial w}{\partial t} = b \tag{1.3'}$$

$$\frac{\partial u}{\partial x} + \frac{\partial v}{\partial y} + \frac{\partial w}{\partial z} = 0 \tag{1.4'}$$

$$\frac{\partial b}{\partial t} = \beta^2 w \tag{1.5'}$$

where $\Phi = p/\rho_0$, $\beta^2 = k/C_p\text{-}N^2$ ($N^2$ is the buoyancy frequency). Here the relation $J/C_p = kw$ has been used in equation (1.5). Note that $\Phi$ here is not really a geopotential field as defined in meteorology but it can represent well the geopotential field. A special attention should be mentioned is that (1.3') is a partial equation with time only. It is the equation for the vertical motion at one point (x,y,z) and its solution, therefore, is the change of vertical motion *at one point*, not following an upward current. The development of the vertical motion can be imagined as in the figure 1, which will be very different if we have a total derivative. This profile for vertical motion as in fig. 1 is typical for any vertical motion in the atmosphere because the impenetrable surface and the upper stable stratosphere will force vertical motion to be zero at these boundaries. From (1.3') and (1.5'), we get

$$\frac{\partial^2 w}{\partial t^2} = \beta^2 w \tag{1.6}$$

Solution of (1.6) clearly is:

$$w = W_1(x,y,z)e^{\beta t} + W_2(x,y,z), \tag{1.7}$$

where $W_1$ is the initial vertical velocity, $W_2$ is a constant distribution of the vertical velocity which is not effected by a positive feedback and will be neglected. Take the partial derivative of (1.2') with respect to time and using (1.1') and rearrange:

$$(\frac{\partial^2}{\partial t} + f^2)v = -\frac{\partial^2 \phi}{\partial y \partial t} + f \frac{\partial \phi}{\partial x} \tag{1.8}$$

Take partial derivative of (1.4') and using (1.1'):

$$(f \frac{\partial}{\partial x} + \frac{\partial^2}{\partial t \partial y})v = \frac{\partial^2 \phi}{\partial x^2} - \frac{\partial^2 w}{\partial t \partial z} \tag{1.9}$$

Take $(f \frac{\partial}{\partial x} + \frac{\partial^2}{\partial t \partial y})$ (1.8) - $(\frac{\partial^2}{\partial t} + f^2)$ (1.9) and rearrange:

$$(f \frac{\partial}{\partial x} + \frac{\partial^2}{\partial t \partial y})(f \frac{\partial}{\partial x} - \frac{\partial^2}{\partial t \partial y})\phi = (\frac{\partial^2}{\partial t^2} + f^2)(\frac{\partial^2 \phi}{\partial x^2} - \frac{\partial^2 w}{\partial t \partial z}) \tag{1.10}$$

Expand and rearrange (1.10), we finally get:

$$\frac{\partial}{\partial t^2}(\frac{\partial^2}{\partial x^2} + \frac{\partial^2}{\partial y^2})\phi = (\frac{\partial^2}{\partial t^2} + f^2)\frac{\partial^2 w}{\partial z \partial t} \tag{1.11}$$

Using the explicit expression for w, (1.7), and plug into (1.11) lead to:

$$\frac{\partial}{\partial t^2}(\frac{\partial^2}{\partial x^2} + \frac{\partial^2}{\partial y^2})\phi = (\beta^2 + f^2)\beta \frac{\partial W_1}{\partial z} e^{\beta \cdot t} \tag{1.12}$$

From (1.12), we get a solution of the form

$$(\frac{\partial^2}{\partial x^2} + \frac{\partial^2}{\partial y^2})\phi = \frac{(\beta^2 + f^2)}{\beta} \frac{\partial W_1}{\partial z} e^{\beta \cdot t} \tag{1.13}$$

It should be noted that solution (1.13) is just a particular solution. (1.12) has many other solutions. However, we are here considering the early development of hurricanes and thus expect that the final solution will grow with time. All other solutions, which increase linearly with time or additional constants, are neglected when compared to the exponential increase.

In order to solve for variable Φ, it is essential to know the functional form $W_1$. We will assume that $W_1$ is a separable function of z and (x,y): $W_1(x,y,z)=H(z)F(x,y)$. Let us consider two cases:

a. *F(x,y) is Dirac-delta function: $F(x,y)=\delta(x_o,y_o)$. This case corresponds to a strong pulse of vertical motion a the point $(x_o,y_o)$.*

In this case, (1.13) becomes:

$$(\frac{\partial^2}{\partial x^2}+\frac{\partial^2}{\partial y^2})\phi = \frac{(\beta^2+f^2)}{\beta}\frac{dH}{dz}e^{\beta \cdot t}\delta(x_0,y_0) \qquad (1.14)$$

(1.14) is an equation for 2D Green function, so its solution is:

$$\phi(x,y,z) = -D\frac{(\beta^2+f^2)}{2\pi\beta}\frac{dH}{dz}e^{\beta \cdot t}\ln\frac{1}{r}+\phi_0(t,z) \qquad (1.15)$$

Where D is a dimensional number of unit magnitude $D=1(m^2)$, so that the function Φ has a correct dimension. This number will be written explicitly so that the final results are always consistent in dimension. Because solution (1.15) is not a harmonic function at infinity, we need to confine (1.15) within a closed region of maximum radius $R_m$ (which can be made as large as needed but it must be finite). Outside this radius, Φ is constant and equal to $Φ_0$. By imposing this condition, (1.15) has the final form as:

$$\phi(x,y,z) = \phi_0 - D\frac{(\beta^2+f^2)}{2\pi\beta}\frac{dH}{dz}e^{\beta \cdot t}\ln\frac{R_m}{r} \qquad (1.16)$$

We now compare this result (fig. 2a) with an observation from a real hurricane (fig. 2b). Fig 2b is the observation of the geopotential field at the 500mb surface of hurricane Anita at its mature stage. It may be ambiguous when comparing fig 2a, which is the solution at the early stage of hurricanes, with fig 2b, which is the observation at the mature stage. However, it should be noticed that solution (1.16) is a separable function of time and spatial coordinates. Therefore, (1.16) will preserve its dependence on radius with time. The only change is the magnitude of this function. As a result, if the theoretical geopotential field of a hurricane has a shape as shown in fig.2a at the initial time, it will continue to be so at the later time. Having said so, it is enough to compare the shape of a theoretical solution at the initial time with observations at the mature stage of a hurricane. This argument will also be applied to later verifications of theoretical solutions with observations.

When the radius approaches zero, solution (1.16) gives a wrong description of geopotential field. This is due to the assumption of Dirac-delta function of vertical velocity. In a real atmosphere, an upward region must have some finite dimension. However, the quite well-fit shape at the large radius tells us that any modification may be expected to contain a natural logarithm expression of radius at the large distance. Because at this moment we still don't have the correct functional form of the function $\Phi$ yet, the calculation of wind fields will be postponed until a correct function for $\Phi$ is obtained. Consider now the second case:

b. *F(x,y) is a top-hat function: F(x,y)=1 for r-$r_o$<**a** and zero otherwise. This case corresponds to a strong pulse of the vertical motion around point ($x_o$,$y_o$). However, in this case the dimension of the upward region has a specific dimension rather than zero as in the case of delta function*

With this hypothetical case, solution (1.13) will be applied for each region separately

$$(\frac{\partial^2}{\partial x^2}+\frac{\partial^2}{\partial y^2})\phi = \frac{(\beta^2+f^2)}{\beta}\frac{dH}{dz}e^{\beta.t} \qquad \text{for r} \leq \mathbf{a} \qquad (1.17a)$$

$$\left(\frac{\partial^2}{\partial x^2}+\frac{\partial^2}{\partial y^2}\right)\phi = 0 \qquad \text{for } r > a \qquad (1.17b)$$

Equations (1.17a,b) will be solved for each region separately and the continuous conditions at r = **a** will be used to match the solutions between the two regions. Because of the symmetry of problem around point *(x_o,y_o)*, it is better to write (1.17a,b) in the cylindrical coordinate with the origin at *(x_o,y_o)*: (1.17a) will become:

$$\frac{1}{r}\frac{d}{dr}\left(r\frac{d\phi_1}{dr}\right) = \frac{(\beta^2 + f^2)}{\beta}\frac{dH}{dz}e^{\beta.t} \qquad (1.18)$$

Since the RHS of (1.18) is a function z only, it is possible to integrate both sides of (1.18) with respect to radius and get the following solution:

$$\phi_1 = \frac{(\beta^2 + f^2)}{\beta}\frac{dH}{dz}e^{\beta.t}\frac{r^2}{4} + C_1 \qquad (1.19a)$$

Where $C_1$ is any constant and will be specified later. Following exactly the same way as above for region II where r>**a**, we obtain:

$$\phi_2 = C_2 - C_3 \ln\left(\frac{1}{r}\right) \qquad (1.19b)$$

Now using the boundary condition that at r>$R_m$, $\Phi$ will be a constant value and is equal to $\Phi_o$. Also, because of the continuous requirement of the function $\Phi$ up to the first derivative at r = **a**, it is simple to specify all three constants $C_1$, $C_2$, $C_3$ and they are equal to:

$$C_1 = \Phi_0 - \frac{(\beta^2 + f^2)}{\beta}\frac{dH}{dz}e^{\beta.t}\frac{a^2}{2}\left(\ln\frac{R_m}{a} + \frac{1}{2}\right)$$

$$C_2 = \Phi_0 \qquad (1.20)$$

$$C_3 = \frac{(\beta^2 + f^2)}{\beta}\frac{dH}{dz}e^{\beta.t}\frac{a^2}{2}$$

The finals solution for function $\Phi$ is:

$$\phi_1 = \Phi_0 + \frac{(\beta^2 + f^2)}{\beta}\frac{dH}{dz}e^{\beta \cdot t}\frac{r^2}{4} - \frac{(\beta^2 + f^2)}{\beta}\frac{dH}{dz}e^{\beta \cdot t}\frac{a^2}{2}(\ln\frac{R_m}{a} + \frac{1}{2}) \quad \text{for } r\leq a \quad (1.19a)$$

$$\phi_2 = \Phi_0 - \frac{(\beta^2 + f^2)}{\beta}\frac{dH}{dz}e^{\beta \cdot t}\frac{a^2}{2}\ln\frac{R_m}{r} \quad \text{for } r > a \quad (1.19b)$$

These theoretical solutions are plotted in fig.3 to compare with the observations of the geopotential field in fig. 2b. It is first realized that the theoretical solutions (1.19a,b) have a U-shape as observed in fig 2b. However, if we pay a little deeper attention to our theoretical solutions, it is then easy to find that no matter how the parameters in (1.19a,b) are changed, the shape of the theoretical solution is still somewhat shallower than observed. The observed geopotential field in fig 2b approaches a constant value very quickly from the core to nearly 50km and then reaches this constant value very slowly. On the contrary, the theoretical curve in fig.3 reaches the constant value much faster than observation (more tilted outside the core region). This subtle difference will be resolved in the non-linear theory in the next section. Once again, as seen in solutions (1.19a,b), both of these solutions increase exponentially with time. Therefore if at the initial time, the U-shape is shallow, it will continue to be so at the later time. The shallowness does not change with time and the difference between the theoretical solutions and the observations still exists. Because of the quite good fit between the analytical solutions and observations, we now attempt to find the wind fields. Plugging (1.19a,b) into (1.8) leads to:

Region I

$$(\frac{\partial^2}{\partial t^2} + f^2)v_1 = -\frac{\beta^2 + f^2}{\beta}e^{\beta \cdot t}\beta\frac{dH}{dz}\frac{y - y_0}{2} + \frac{\beta^2 + f^2}{\beta}e^{\beta \cdot t}f\frac{dH}{dz}\frac{x - x_0}{2} \quad (1.22)$$

by setting $r = \sqrt{(x - x_0)^2 + (y - y_0)^2}$ $\cos\varphi_0 = \beta/\sqrt{(\beta^2 + f^2)}$ and $\sin\varphi_0 = f/\sqrt{(\beta^2 + f^2)}$, (1.22) can re-written as

$$(\frac{\partial^2}{\partial t^2}+f^2)v_1 = -\frac{(\beta^2+f^2)^{3/2}}{2\beta}e^{\beta.t}\frac{dH}{dz}\sin(\varphi-\varphi_0)r \qquad (1.23)$$

From (1.23), it is simple to obtain solution for v-component (note that this is v-component in Cartesian coordinate, not in cylindrical coordinate).

$$v_1 = -\frac{(\beta^2+f^2)^{1/2}}{2\beta}e^{\beta.t}\frac{dH}{dz}\sin(\varphi-\varphi_0)r \qquad (1.24a)$$

If (1.1') and (1.2') are combined, It is easy to get an equation for u-component as:

$$(\frac{\partial^2}{\partial t^2}+f^2)u = -\frac{\partial^2\phi}{\partial t\partial x} - f\frac{\partial\phi}{\partial y} \qquad (1.25)$$

After following the same way as for v-component, a solution for u-component in region I will be given by:

$$u_1 = -\frac{(\beta^2+f^2)^{1/2}}{2\beta}e^{\beta.t}\frac{dH}{dz}\cos(\varphi-\varphi_0)r \qquad (1.26a)$$

Region II

Substitute solution (1.19b) to equations (1.8) and (1.25), after some manipulations the solutions of the u and v components in region II are:

$$v_2 = -\frac{(\beta^2+f^2)^{1/2}}{2\beta}e^{\beta.t}\frac{dH}{dz}\sin(\varphi-\varphi_0)\frac{a^2}{r} \qquad (1.24b)$$

$$u_2 = -\frac{(\beta^2+f^2)^{1/2}}{2\beta}e^{\beta.t}\frac{dH}{dz}\cos(\varphi-\varphi_0)\frac{a^2}{r} \qquad (1.26b)$$

If we plot solution (1.24a,b) and (1.26a,b), it can be verified easily that these solutions are indeed for a cyclonic motion and the total wind $V=\sqrt{u^2+v^2}$ has a profile fit well with observation as shown in fig.4 and fig. 2b. Note that in the case vertical wind decreases with height (in the upper half of the atmosphere: where dH/dz<0), this profile is unchanged.

However, the tangential wind will now immediately become anti-cyclonic in both region I and II as seen in solutions (1.24a,b and 1.26a,b) (It is not very easy to see this conclusion directly in the Cartesian coordinate. We need to plot them out explicitly). It is very interesting that in a numerical study, Moller and Shapiro (2002) have pointed out that a hurricane does have a very large anti-cyclonic region throughout the upper half of the atmosphere far outside the core region. But anyway, the tangential wind near the core region is still cyclonic in their study and the linear theory is still inaccurate.

Summary:

Strengths: The tangential wind profile has a correct behavior with radius as observed. It first increases linearly with radius and then decreases as a function of inverse radius. The cyclonic sense of wind is also captured *whenever vertical velocity increases with height* (dH/dz>0), which is correct in the lower half of a real hurricane. Both the tangential wind and radial wind increases as an exponential function of time, which tell us about the growing of hurricanes as expected. The theoretical geopotential field is given by a parabolic shape in the region where the tangential wind increases linearly with radius. This analytical curve then approaches logarithmically to a constant value at the large radius, which also fit pretty well observations. Vertical wind increases uniformly as an exponential function of time within the region where positive feedback is effective and zero outside.

Weaknesses: This linear theory will immediately give a hill shape of gepotential and anticyclonic wind whenever vertical motion decreases with height (violate 1A, 1B). This is not true because, in the upper half of a real hurricane where the vertical wind decreases with height, the U shape of geopotential still predominant and the tangential wind is still cyclonic up to 13km It is interesting to note that above 13km, the tangential wind is anticyclonic even near the core region as seen in fig. 5b. That the linear theory gives a hill-shape of geopotential and an anti-cyclonic wind from the level where vertical velocity decreases with height is a proof of the useless of the 2D Rankine vortex in describing hurricanes. If we work out a little more, it is straightforward to find that the vertical profile of tangential wind also does not really fit observation and thus violate observation 1E. In addition, because of confining ourselves to the early stage of hurricane development, this theory is expected to be correct up to first 6-12h of growing. Beyond this period, this theory will give inaccurate descriptions (unlimited growing).

**3.2. Nonlinear theory**

In this section, the linear system in section 3.1 will be generalized so that nonlinear terms are included. The simple idealized theory in section 3.1 will be very helpful to us in obtaining the solutions of the following nonlinear theory as we shall see. It is expected that the nonlinear theory can explain the shallowness of the theoretical geopotential curve obtained in the linear theory and, more importantly, it must explain for the U-shape of geopotential as well as the cyclonic wind of the tangential wind in the upper half of the atmosphere, where wind is divergent. As mentioned in section 3.1, it was not possible to explain these facts with the linear theory. Now, we make some assumptions for a new nonlinear theory for the early stage of hurricane development. The meaning of each assumption was explained in section 3.1.

Assumptions:

- Non-linear terms in the tangential and radial momentum equations are included, but non-linear terms *in vertical and thermodynamic equations* are neglected.
- No friction is considered

- Incompressible fluid dρ/dt=0
- Boussisnesq approximation
- There exist a region of radius "a" around the point $(x_o, y_o)$ in which vertical velocity (w) is related to heating source (J) by a positive feedback relationship (J=kw)
- F-plane
- Hurricanes are axis- symmetric

To facilitate our consideration in this case, it is better to start the system of primitive equations in the cylindrical coordinate:

$$\frac{\partial u}{\partial t} + u\frac{\partial u}{\partial r} + v\frac{1}{r}\frac{\partial u}{\partial \lambda} + w\frac{\partial u}{\partial z} - \frac{v^2}{r} = -\frac{\partial \phi}{\partial r} + fv \quad (2.1)$$

$$\frac{\partial v}{\partial t} + u\frac{\partial v}{\partial r} + v\frac{1}{r}\frac{\partial v}{\partial \lambda} + w\frac{\partial v}{\partial z} + \frac{uv}{r} = -\frac{1}{r}\frac{\partial \phi}{\partial \lambda} - fu \quad (2.2)$$

$$\frac{\partial w}{\partial t} = b \quad (2.3)$$

$$\frac{1}{r}\frac{\partial}{\partial r}(ru) + \frac{1}{r}\frac{\partial v}{\partial \lambda} + \frac{\partial w}{\partial z} = 0 \quad (2.4)$$

$$\frac{\partial b}{\partial t} = J \quad (2.5)$$

Because of the symmetry of hurricanes with respect to the azimuthal angle, the system ((2.1) to (2.5)) now becomes simpler:

$$\frac{\partial u}{\partial t} + u\frac{\partial u}{\partial r} + w\frac{\partial u}{\partial z} - \frac{v^2}{r} = -\frac{\partial \phi}{\partial r} + fv \quad (2.1')$$

$$\frac{\partial v}{\partial t} + u\frac{\partial v}{\partial r} + w\frac{\partial v}{\partial z} + \frac{uv}{r} = -fu \quad (2.2')$$

$$\frac{\partial w}{\partial t} = b \quad (2.3')$$

$$\frac{1}{r}\frac{\partial}{\partial r}(ru) + \frac{\partial w}{\partial z} = 0 \qquad (2.4')$$

$$\frac{\partial b}{\partial t} = J \qquad (2.5')$$

Our purpose now is to solve this new system. As in section 3.1, combination of (2.3') and (2.5') gives us a solution for w-component:

$$W(r,z,t) = W_1(r,z)e^{\beta \cdot t} + W_2(r,t) \qquad (2.6)$$

Because of the reasonable consistence of the assumption that $W_1(x,y,z)$ is expressed as product of a top-hat function of (x,y) and a function of z: $W_1(x,y,z) = H(z)F(x,y)$ as seen in section 3.1b, we will utilize this separable function from the beginning and solve the system ((2.1') to (2.5')) in each region I and II separately.

Region I

From (2.4') and using $W_1(x,y,z) = H(z)$ (since F(x,y)=1 in this region), we have

$$\frac{1}{r}\frac{\partial}{\partial r}(ru) = -\frac{\partial w}{\partial z} = -\frac{dH}{dz}e^{\beta \cdot t} \qquad (2.7)$$

Here $W_2(x,y)$ was set to zero as in section 3.1 because all of the vertical motions in region I are assumed to originate from the positive feedback process. Integrate (2.7) with respect to radius, a solution for the radial wind will be given by:

$$u_1 = -\frac{dH}{dz}e^{\beta \cdot t}\frac{r}{2} + \frac{C(z,t)}{r} \qquad (2.8)$$

Since at r=0, it is obvious that there is no u-component, so C(z,t) = 0. Plugging (2.6) and (2.8) into (2.2') to obtain an equation for the tangential wind in region I:

$$\frac{\partial v}{\partial t} + (-\frac{dH}{dz}e^{\beta \cdot t}\frac{r}{2})\frac{\partial v}{\partial r} + He^{\beta \cdot t}\frac{\partial v}{\partial z} + (-\frac{dH}{dz}e^{\beta \cdot t}\frac{r}{2})\frac{v}{r} = -f(-\frac{dH}{dz}e^{\beta \cdot t}\frac{r}{2}) \qquad (2.9)$$

From this equation, it is clear that the tangential wind is a function of (r, z, t). It is very difficult to solve for this equation. However, the simplified linear theory in section 3.1b suggests that v may be a linear function of radius. So we attempt to find the tangential wind in the form of: v = F(z,t)r and substitute this into (2.9). Also, to avoid unnecessary cumbersome symbol, set $Q = -\frac{1}{2}\frac{dH}{dz}$, (2.9) now becomes:

$$r\frac{\partial F}{\partial t} + rQe^{\beta t}F + rHe^{\beta t}\frac{\partial F}{\partial z} + rQe^{\beta t}\frac{Fr}{r} = -rfQe^{\beta t} \quad (2.10)$$

Since this equation must hold for every r, z, t, it must satisfy:

$$\frac{\partial F}{\partial t} + Qe^{\beta t}F + He^{\beta t}\frac{\partial F}{\partial z} + Qe^{\beta t}F = -fQe^{\beta t} \quad (2.11)$$

or

$$\frac{\partial F}{\partial t} = -e^{\beta t}(H\frac{\partial F}{\partial z} + 2QF + fQ) \quad (2.12)$$

Consider first the homogenous solution of (2.12):

$$\frac{\partial F_m}{\partial t} = -e^{\beta t}(H\frac{\partial F_m}{\partial z} + 2QF_m) \quad (2.13)$$

Up until this point nothing about the functional form of H(z), which is the profile of vertical velocity with height, has been mentioned. The appearance of the upper tropopause makes problem become special. No matter how large the buoyancy is, this tropopause (a kind of upper rigid lid) will make vertical velocity become zero right below it. The impact of this lid in some way can be thought of as an external forcing of an external mechanism and so far can not be described by the system ((2.1') to (2.5')). Because the functional form of the vertical motion (H(z) in (2.7)) is not derived from the system ((2.1') to (2.5')) and somewhat arbitrary, the effect of boundaries can be included by choosing a prescribed function such that the vertical velocity will be zero at the tropopause and surface. It is reasonable to choose: H(z)=$W_0$sin($\lambda$z), where $\lambda$ is expected to be inversely proportional to the standard height $H_0$ ($H_0 \approx$10km) so that vertical velocity is equal to zero at z=0 and z=$H_0$. This choice, however, is not unique. Several functional forms can be tested, such as the parabolic function, but the behavior of final solutions should be the same because the system ((2.1') to (2.5')) is the first-order partial differential system. From the definition of $Q = -\frac{1}{2}\frac{dH}{dz}$, we have: Q= -$W_0\lambda$cos($\lambda$z)/2. Plug these expressions of H and Q into (2.13):

$$\frac{\partial F_m}{\partial t} = -e^{\beta \cdot t}(W_0 \sin(\lambda z)\frac{\partial F_m}{\partial z} - W_0 \lambda \cos(\lambda z) F_m) \qquad (2.14)$$

The form of this equation suggests us that a growing solution will have the form:

$$F_m = \exp(\mu e^{\beta \cdot t}) G(z) \qquad (2.15)$$

where $\mu$ is an arbitrary positive number. Substitute this solution (2.15) into (2.14):

$$\mu \exp(e^{\beta \cdot t}) e^{\beta \cdot t} \beta G = -e^{\beta \cdot t}(W_0 \sin(\lambda z)\frac{dG}{dz} - W_0 \lambda \cos(\lambda z) G)\exp(e^{\beta \cdot t}) \qquad (2.16)$$

or $$G\mu\beta = -(W_0 \sin(\lambda z)\frac{dG}{dz} - W_0 \lambda \cos(\lambda z) G) \qquad (2.17)$$

$$\Leftrightarrow \frac{dG}{G} = \frac{(W_0 \lambda \cos(\lambda z) - \mu\beta)}{W_0 \sin(\lambda z)} dz \qquad (2.18)$$

Integrate (2.18) with respect to z:

$$\ln(G) = \ln(\sin(\lambda z)) - \frac{\mu\beta}{W_0 \lambda} \ln(\tan(\frac{\lambda z}{2})) \qquad (2.19)$$

And finally, the solution for function G(z) is:

$$G = G_0 \frac{\sin(\lambda z)}{\{\tan(\frac{\lambda z}{2})\}^{\frac{\mu\beta}{W_0 \lambda}}} \qquad (2.20)$$

From (2.15), we thus obtain a homogeneous solution for function $F_m$.

$$F_m = G_0 \frac{\sin(\lambda z)}{\{\tan(\frac{\lambda z}{2})\}^{\frac{\mu\beta}{W_0 \lambda}}} epx(\mu e^{\beta \cdot t}) \qquad (2.21)$$

By inspecting (2.14) again, it is not hard to find another separable solution:

$$F_m' = D\sin(\lambda z)$$

Where D is any constant with a correct dimension. So the homogeneous solution of (2.14) finally is

$$F_m = G_0 \frac{\sin(\lambda z)}{\{\tan(\frac{\lambda z}{2})\}^{\frac{\mu\beta}{W_0 \lambda}}} epx(\mu e^{\beta \cdot t}) + D\sin(\lambda z) \qquad (2.21')$$

Note that (2.14) has an infinitive number of separable solutions but we are here just considering the largest growing solution with time which is finite at z = 0. The requirement of finiteness of solution (2.21') at $z = 0$ puts a restriction on values of $\mu$, $\beta$, $\lambda$ and $W_0$. By taking a limitation of (2.21), it is then immediate to obtain that:

$$\mu\beta/(\lambda W_0) < 1$$

Given $\beta$, $\lambda$ and $W_0$, the largest possible of parameter $\mu$ is: $\mu_{max} = \lambda W_0/\beta$. This maximum value of $\mu$ will result in a solution growing fastest with time and we will consider this solution only. We will choose $\mu \approx \lambda W_0/\beta$ in (2.21') and understand implicitly that this choice will always guarantee $\mu\beta/(\lambda W_0) < 1$ so that solution (2.21') is finite at $z = 0$. It is usual to find the final solution of an inhomogeneous equation by setting $F(z,t)=G(z,t)F_m(z,t)$ and plug it into (2.12). By following this way, we will come up with an even more complicated than the original equation. However, it is possible to eliminate the non-homogenous term by realizing that the last two terms in the RHS of (2.12) are linear in Q and contain no derivative in F, so a particular solution of (2.12) can be chosen as follows:

$$F_p = -\frac{f}{2} \tag{2.22}$$

The final solution for tangential wind in region I is:

$$v_1 = (F_m + F_p)r = \{G_0 \frac{\sin(\lambda z)}{\left(\tan(\frac{\lambda z}{2})\right)^{1-\delta}} epx(\frac{W_0 \lambda}{\beta} e^{\beta t}) + D\sin(\lambda z) - \frac{f}{2}\}r \tag{2.23}$$

where $\delta$ is a very small positive number (but must be different from zero). The profile of the homogeneous solution $F_m$ is plotted in fig. 5a to see how it varies with height since it will tell us about the vertical profile of tangential wind. (Note that in the linear theory in section 3.1, the profile for tangential wind is a cosine function of height if $H(z)=W_0\sin(\lambda z)$. This means that wind will be anticyclonic as soon as there is a divergence of wind (vertical velocity decreases with height dH/dz<0)). As seen in Fig.5a, we now have a tangential wind decrease with height and still have a cyclonic sense when dH/dz < 0 as expected, provided that the homogeneous solution is large enough to overcome the particular solution –f/2. Only at very high altitude does particular solution –f/2 have a possibility to become greater than the homogeneous solution $F_m$ and an anti-cyclonic tangential wind will appear at high altitude only, exactly as observed in fig. 5b.

Several interesting points of solution (2.23) should be recapitulated. First, it gives us a correct description of the tangential wind with radius near the core of a hurricane, a linear increase. Secondly and more interestingly, the tangential wind in this region increases as an exponential function of exponential of time, which is much stronger than radial wind (exponential of time only). This much stronger increase is due to the appearance of the vertical advection terms in equation (2.2'). If these vertical advections are eliminated, the change in tangential wind with time will be exponential of time only, which is the same as that in section 3.1b for the linear theory. The much faster increase of tangential wind with time in the core region may be expected if an eye is formed at the later time. Another feature is that solution (2.23) is a decrease function of height, which means that tangential wind will decrease with height. Finally, the cyclonic motion still exists at divergent levels as long as the first two terms in (2.23) are greater than the last term.

Region II

As in region I, we first solve for the radial wind in this region. Because w = 0 in region II (or at least a function of time and radius only), it is immediate to integrate (2.4') with radius to obtain the solution for the radial wind:

$$u_2 = C_1 - \frac{C_2}{r} \tag{2.25}$$

Using the continuous condition of radial wind at radius r = **a** and zero at infinity (we can regard $R_m$ as a very large radius so that approximately we have a infinitive closed region), it is simple to find that: $C_1=0$ and $C_2 = \frac{1}{2}\frac{dH}{dz}e^{\beta.t}\frac{a^2}{r}$. So the exact solution for the radial wind in region II is

$$u_2 = -\frac{1}{2}\frac{dH}{dz}e^{\beta.t}\frac{a^2}{r} \equiv e^{\beta.t}\frac{Q'}{r} \tag{2.26}$$

Where the symbol $Q' = -\frac{1}{2}\frac{dH}{dz}a^2$ was introduced for abbreviation. Substitute this solution into (2.2') with attention that w = 0, we have

$$\frac{\partial v_2}{\partial t} + u_2\frac{\partial v_2}{\partial r} + w_2\frac{\partial v_2}{\partial z} + \frac{u_2 v_2}{r} = -fu_2 \tag{2.27}$$

$$\Rightarrow \frac{\partial v_2}{\partial t} + \frac{Q'}{r}e^{\beta.t}\frac{\partial v_2}{\partial r} + \frac{Q'}{r^2}e^{\beta.t}v_2 = -f\frac{Q'}{r}e^{\beta.t} \tag{2.28}$$

$$\Leftrightarrow \frac{\partial v_2}{\partial t} = -e^{\beta.t}(\frac{Q'}{r}\frac{\partial v_2}{\partial r} + \frac{Q'}{r^2}v_2 + f\frac{Q'}{r}) \tag{2.29}$$

With the help from the linear system in section 3.1b, we will find solution of (2.29) in the form of: $v_2 = \frac{G(z,t)}{r}$ and substitute it into (2.29):

$$\frac{1}{r}\frac{\partial G}{\partial t} = -e^{\beta.t}(\frac{Q'}{r}(-\frac{G}{r^2}) + \frac{Q'}{r^2}\frac{G}{r} + f\frac{Q'}{r}) \tag{2.30}$$

so

$$\frac{1}{r}\frac{\partial G}{\partial t} = -e^{\beta.t}f\frac{Q'}{r} \tag{2.31}$$

From assumption that H(z)=$W_0 \sin(\lambda z)$, we have Q'= -$a^2 \lambda W_0 \cos(\lambda z)/2$. After substitute Q' into (2.31) with the note that r > 0 in region II, it is straightforward to integrate (2.31) with time directly (Q' does not depend on t) and get:

$$G(z,t) = -e^{\beta t} \frac{fQ'}{\beta} + C(z) = W_0 e^{\beta t} \frac{fa^2 \lambda}{2\beta} \cos(\lambda z) + C \qquad (2.32)$$

Finally, the solution for the tangential wind in region II is:

$$v_2(r,z,t) = \{W_0 e^{\beta t} \frac{fa^2 \lambda}{2\beta} \cos(\lambda z) + C\} \frac{1}{r} \qquad (2.33)$$

Since $\cos(\lambda z)$ is greater than zero for 0<$\lambda z$<$\pi/2$ (the lower half of the atmosphere), the tangential wind in the lower half of the atmosphere will be cyclonic (greater than zero). In the upper half of the atmosphere, $\cos(\lambda z)$ is smaller than zero because $\pi/2$<$\lambda z$< $\pi$. However, it is interesting that motion can still be cyclonic in the upper atmosphere if the constant C(z) is large enough. We thus have a cyclonic motion as expected for the lower atmosphere and even for the upper atmosphere where divergence occurs. This feature could not be obtained by the linear theory in section 3.1. However, there no longer has a continuity of the tangential wind at r = **a** as in the linear theory. This is because, when vertical advection is included, basically, the tangential momentum equation has been added a source term which is zero outside the radius r = **a** and is an exponential function of time inside. Therefore, solution (2.23) for the tangential wind increases very strongly with time while solution (2.33) does not. The hurricane development at the early stage can be imagined according to this non-linear theory as in the fig. 6.

It is very interesting that in the early developing stage of a real hurricane, observations have shown that, at the early phase of hurricane development, there is a very rapid increase of the tangential wind inside the core region and a much slower increase of the tangential wind outside as shown in fig.7 for hurricane DIANA. Clearly this feature is captured very well by solutions (2.23) and (2.33). Many other examples can be found in Willoughby *et. al*. (1990).

The final task is to find a solution for the function Φ, which has the meaning the same as a geopotential field in meteorology. We expect that with this new nonlinear theory, it is then possible to have a U-shape of geopotential when vertical velocity decreases with height as observed. As noted in section 3.1, the linear theory could not possess this feature. Whenever dH/dz<0, according to this linear theory, motion will be anticyclonic and a high pressure system will develop. (In a real hurricane, hurricanes do have anticyclonic motion but it is just near the tropopause, whereas dH/dz is less than zero from the middle.). Starting with equation (2.1')

$$\frac{\partial u}{\partial t} + u\frac{\partial u}{\partial r} + w\frac{\partial u}{\partial z} - \frac{v^2}{r} = -\frac{\partial \phi}{\partial r} + fv \qquad (2.34)$$

In region I, using $v_1$, $w_1$, $u_1$ and plug into (2.34), we have

$$\beta Q e^{\beta t} r + Q e^{\beta t} r Q e^{\beta t} + H e^{\beta t} e^{\beta t} r \frac{dQ}{dz} - K^2 r = -\frac{\partial \phi}{\partial r} + fKr \qquad (2.35)$$

where K = $\{G_0 \frac{\sin(\lambda z)}{\tan(\frac{\lambda z}{2})^{\frac{1}{W_0 \beta \lambda}}} epx(e^{\beta t}) + D\sin(\lambda z) - \frac{f}{2}\}$ and Q = - $W_0\lambda\cos(\lambda z)/2$. Rearrange

(2.35):

$$\frac{\partial \phi}{\partial r} = -\{Q\beta e^{\beta t} + Q^2 e^{2\beta t} + He^{2\beta t}\frac{dQ}{dz} - K^2 - fK\}r \qquad (2.36)$$

From (2.36), we obtain a solution for Φ as:

$$\phi = \Phi_0 - \{K^2 + fK - Q\beta e^{\beta t} - Q^2 e^{2\beta t} - He^{2\beta t}\frac{dQ}{dz}\}\frac{r^2}{2} \qquad (2.37)$$

Because H ≈ sin($\lambda z$), Q ≈ - dH/dz will be less than zero in the lower atmosphere and dQ/dz = -d$^2$H/dz$^2$ is greater than zero. In this case, the first three terms inside the bracket in (2.37) are greater than zero and the last two terms are smaller than zero. Note that since K is a very sensitive function of time, K-terms will dominate at later time and the sum inside the bracket on the RHS of (2.37) will tend to be positive. Therefore, a low pressure system (U-shape of geopotential) will develop at the lower atmosphere.

Now consider the case in which dH/dz<0 as it is in the upper atmosphere. In this case Q>0 and dQ/dz<0. In this case only the first two terms inside the bracket of (2.37) are positive. Note here again that the term K$^2$ is always positive and it is the most weighted term. Therefore, the bracket term in the RHS of (2.37) is still positive and we still have a U-shape of geopotential. These are excellent results which are not obtained by the linear theory. Near the tropopause, K can be smaller than zero because of the strong decrease of K with height (fig. 5a) and there may have an anticyclonic motion exactly as observed. Interesting is that, even in the case K<0, it is still possible to have the U-shape of geopotential.

The final task now is to find an expression for the function Φ in region II by using $u_2$, $v_2$ (remember that we have chosen the vertical velocity to be zero in region II). Plug all these solutions in to equation (2.1'):

$$\beta e^{\beta.t} \frac{Q'}{r} - e^{\beta.t} \frac{Q'}{r} e^{\beta.t} \frac{Q'}{r^2} - \frac{C^2}{r^3} = -\frac{\partial \phi}{\partial r} + f \frac{C}{r} \qquad (2.38)$$

Where C = $\{W_0 e^{\beta.t} \frac{fa^2 \lambda}{2\beta} \cos(\lambda z) + C(z)\}$ and Q' is defined from (2.26), rearrange (2.38) leads to:

$$\frac{\partial \phi}{\partial r} = -\{\beta e^{\beta.t} \frac{Q'}{r} - e^{2.\beta.t} \frac{Q'^2}{r^3} - \frac{C^2}{r^3} - \frac{fC}{r}\} \qquad (2.39)$$

Integrate (2.39) with respect to radius with the note that Q' and C don't depend on the radius, it is easy to obtain a solution for the geopotential in region II:

$$\phi_2 = \Phi_0 - \{\beta e^{\beta \cdot t} Q' \ln r + e^{2\cdot\beta\cdot t}\frac{Q'^2}{r^2} + \frac{C^2}{r^2} - fC \ln r\} \qquad (2.40)$$

If region II is confined within a radius of $R_m$ and beyond that $\Phi$ is equal to a constant value, (2.40) can be rewritten as:

$$\phi = \Phi_0 - \{\beta e^{\beta \cdot t} Q' \ln \frac{R_m}{r} + e^{2\cdot\beta\cdot t}\frac{Q'^2}{r^2} + \frac{C^2}{r^2} - fC \ln \frac{R_m}{r}\} \qquad (2.41)$$

The geopotential field now decreases not only as a natural logarithm of the inverse of radius but a little faster by the terms that are inversely proportional to the square of the radius. These extra terms make the theoretical solution (2.41) fit better with the observed curve in fig. 2b. This little change explains for the deep U-shape of the observation of geopotential in fig 2b, which can not be accounted for in the linear theory. In the non-linear theory, the geopotential field inside region I will be deepened much faster (given by (2.40)) than that in region II (given by (2.41)) and we thus have once again the discontinuity of geopotential at later time.

Summary:

Strengths: This non-linear theory so far captures all the good points in the linear theory (as it must) such as the linear increase of the tangential wind with radius and subsequent decrease as a function of inverse radius, the cyclonic sense of wind, the U-shape of geopotential field and so on. In addition, this theory now offers some new properties. First, the tangential wind in the region where the positive feedback is effective will no longer increase as an exponential of time. Instead, it increases much stronger, as an exponential function of exponential of time ($\exp(e^{kt})$), which is much faster than the radial wind or the tangential wind outside this feedback region. This may help us explain why hurricanes have an eye formed in this region (it does not mean that hurricanes have an eye at this moment. It just means that the tangential wind tends to form an eye). Secondly, this nonlinear theory gives us a decrease of tangential wind with height, which is exactly what we observed. Finally, this theory is also able to give us a potentially deeper U-shape of geopotential as well as the cyclonic motion even when there is a divergence in the upper half of the atmosphere, which was not obtained by the linear theory.

Weaknesses: In this nonlinear theory, we now cannot expect to have continuous tangential wind and geopotential fields. This is because we have assumed a top-hat function of the feedback mechanism. This top-hat function creates an exponential increase with time of vertical wind and, consequently, the vertical advection within the region subjected to positive feedback. Outside this region, vertical velocity is zero and thus vertical advection is zero too. The root of this discontinuity lies in the top-hat function assumption and will be removed Part II. Also, as in the linear theory, because we confined ourselves to the early stage of hurricane development, this theory is also expected to be correct up to first 6-12h of growing. Beyond this period, this theory may give inaccurate descriptions.

## 4. Unresolved problems

First, as we can find throughout this work, we always assume the existence of a positive feedback between latent heat source and vertical motion. Now the question is *under which conditions is this feedback effective?* Clearly, there always has vertical motion in the atmosphere but hurricanes appear rarely. This work basically, so far, gives no information about when this feedback mechanism will appear. Instead, it has been assumed implicitly that this positive feedback will always exist so that a self-induced system will develop. Once this assumption is meaningful, we will have the solutions that have been found in sections 3.1 and 3.2. This positive feedback is very vital for these above theories. However, it should be linked in some way with friction so that the feedback will be controlled. Another feedback mechanism will be considered in an upcoming paper.

Second, in both the linear and non-linear theories, wind and geopotential field will increase with time endlessly. This is not true because when hurricanes reach a mature stage, another process will appear or dominate to consume/limit the source of energy so that hurricanes can not grow forever. The root of the unlimited growing lies once again in the assumption of the positive feedback. In all of the above theories, this feedback is independent from any internal mechanism so that it is always possible to find a solution for radial wind and vertical wind independently. Using these radial and vertical winds, we then find tangential wind and geopotential field. Basically, the system of primitive equations has been separated into two sub-systems: one for the radial and vertical wind and one for the tangential wind and geopotential. This independence will be removed if we can find a way to link a feedback process to the tangential wind.

Third, we have neglected all the advection terms in the vertical momentum and thermodynamic equations so that the closed system between radial and vertical wind ((2.3') and (2.5')) is simplified. This assumption, however, is not a major weakness because hurricanes are azimuthally symmetric, and we will always have a closed system between vertical wind and radial wind as long as the positive feedback is related to radial and vertical wind only. Normally, these advection terms are largest near the surface or the tropopause because radial wind is strongest here. The inclusion of the advection terms in the vertical momentum equation and thermodynamic equation will make problem become very complicated. However, it is fortunate that, at these levels, the radial gradient of both vertical wind and temperature ($\partial w/\partial r$, $\partial b/\partial r$) are very small. Therefore, these total advections are quite small and it is reasonable to expect that these advection terms will not have any significant impact on the hurricane development.

Finally, *both the linear and non-linear theories in this work do require a pre-existing vortex so that the solutions are valuable*. We can see this point quickly by setting t=0 in all solutions obtained so far, such as (2.6), (2.7), (2.23), (2.33)….Now the question is "does this work really describe the early stage of hurricane development as stated", since it also needs an initial vortex to start with as the traditional approaches do? The root of this requirement lies in two facts: the positive feedback assumption and the applicability of the governing equations. First we should ask ourselves: "Is a positive feedback effective in a quiet atmosphere?" Of course it is not. In order to be effective, this feedback mechanism requires an initial vertical velocity so that latent heat can be released. Therefore, right at the initial time, vertical velocity must be different from zero. Due to the existence of the tropopause and the impenetrable lower surface, vertical velocity will be zero at both these levels. Thus, dw/dz will be different from zero and there is an unavoidable convergence/divergence. If the continuity equation is valid at this initial time, radial wind *must* be different from zero. The other two constraints imposed by equations (2.1') and (2.3'), similarly, cause the tangential wind and geopotential field to have some initial values and the existence of an initial vortex is necessary. We have no way to escape an initial vortex in above theories. Usually in the atmosphere, at the early stage of any meteorological phenomenon, there will appear a transient period in which *no equation can be applied* because this transient period is stochastic and chaotic. Only after some moment, when the phenomenon becomes governed by the fluid equations, can we employ these equations to this phenomenon. Forcing all field variables to grow controllably by the governing equations is equivalent to an assimilation process in a numerical model. The data assimilation can be regarded as a process which forces all field variables to proceed in a balanced way at the initial time. If somehow the atmosphere has vertical motion but radial wind is still zero everywhere, this will immediately mean that the fluid equations are not appropriate yet (the appearance of vertical motion instantly implies that the divergence or convergence will be different from zero because of the upper and bottom boundaries). On the contrary, if we suppose that the governing equations are applicable, by assuming a positive feedback, we automatically have radial wind and geopotential fields different from zero at the initial time. Obviously, no hurricane can develop from a quiet atmosphere. Therefore, a pre-existing vortex is a natural requirement for these above theories. Note that we can, in principal, create this initial vortex nearly as weak as we want, depending on how small the initial vertical velocity is but it can not be zero. It should be careful that the assumption of positive feedback will require that the initial vertical velocity be large enough and the initial vortex thus can not be too weak.

So far, nothing about the eye of a hurricane has been mentioned in this work. The question is "does a hurricane have an eye at the early stage of its development?" Actually, as we can find in any theories from section 3.1 and 3.2, vertical motion is given by an exponential function of time inside the whole region I and zero elsewhere. Is this a really weakness of these theories? I don't think so. A hurricane, at its very first stage, needs to have a very strong mechanism to develop and in this early stage of development, it would not show its eye clearly (Palmen and Newton 1969). It is reasonable to believe that our solutions will not need to contain an eye in the early stage of hurricane development. Rather, these solutions need to have the potential of building an eye as hurricane develop and this is exactly what we obtained in section 3.2 (the tangential wind in region I grows much faster than in region II, (2.23) vs. (2.33))

It is perhaps ambiguous that, instead of introducing a prescribed radius of maximum wind speed as is usually encountered in literatures (Smith 1980; Kuo et. al. 1997), I have launched a region of radius "a" to consider. These above theories appear to employ a scaling parameter of mature hurricanes and, therefore, they are invalid for the early description of hurricane development. However, the point is that I introduced here *not a number* (radius of the positive feedback region) *but a mechanism* (positive feedback) and everything emerges naturally. This is important since we can postulate different tentative mechanisms to investigate the behaviors of the solution analytically. My assumption here about a feedback mechanism leads to a correction in many respects with observations at the early stage of a hurricane, and this is new.

## 5. Discussion and conclusion

By assuming that there exists a positive feedback within a region of radius r = **a** and using first the simple linear theory (section 3.1), we have obtained analytical solutions that have several characteristics that agree with hurricane observations such as the U-shape of geopotential, the linear increase with radius near a core region and the decrease as a function of inverse radius outside this region of tangential wind. These solutions are somewhat the same as those of the 2D Rankine vortex. However, several important aspects of the linear theory are not represented properly. First, the tangential wind does not have a correct profile with height as observed. The tangential wind, according to this linear theory, will be anti-cyclonic when vertical wind decreases with height (divergent). In a real hurricane, observations actually show that cyclonic wind penetrates deeply nearly up to the tropopause and then turns anti-cyclonic in a very shallow layer right below the tropopause (fig.5b). Second, the theoretical geopotential field will immediately have a hill shape in the upper half of the atmosphere where vertical wind decreases with height. In fact, like tangential wind, observations also pointed out that the geopotential field in real hurricanes has a U-shape up to the tropopause regardless of divergence or convergence of wind. Moreover, the U-shape of the theoretical geopotential field is somewhat shallower than that of a real hurricane. Note once again that both the geopotential fields inside and outside the feedback region increase exponentially with time (eqns (1.19a,b)). Therefore, if at the initial time, the U-shape is shallow, it will continue to be so at the later time and the difference between the theoretical solutions and observations still exists. These contradictions show us clearly the weaknesses of the linear theory.

It should be mentioned that these weaknesses do not diminish the role of the simplified linear theory. This theory turns out to be very helpful to us in obtaining the analytical solutions in the non-linear theory as we saw in section 3.2. When a new problem is explored, it is usual to start with a simple problem first. This simplified problem will help us obtain some insights into what is correct, what is not and why it is not correct. This is particularly helpful since, from this information, it is possible to have some good speculations and understandings to make further improvements.

To overcome all of the shortcomings of the linear theory, the full non-linear system of primitive equations has been used (except the vertical momentum and thermodynamic equations). From this theory, we have obtained analytical solutions that fit well with the observations in many respects. In this non-linear theory, a particular technique has been utilized to find the solutions which are separable in the radius, time and height. These are not the only solutions we can get since the solutions obtained so far belong to a separable class and there must exist many other solutions (as many as phenomena we meet in the atmosphere everyday). In different situations, it admits a different set of solutions that describes the situation under consideration. It is a beautiful property of mathematics to contain so many different solutions to a system of equations, but it is even more amazing how the Nature can show us some of these solutions in the physical world. Other unknown solutions may not have physical meaning or they can not express themselves in nature because of some restricted conditions. The solutions obtained so far can be thought of as a solution at the initial phase of a hurricane in the special conditions that I imposed (that is, the existence of a positive feedback region of radius "a"). We may choose other assumptions but after all I believe that it is possible to find an analytical solution for hurricanes satisfactorily. The small-scale processes or the asymmetric properties contribute to the minor characteristics of hurricanes only and will be altered in different cases, but the main behaviors should be the same.

The non-linear theory presented in this work has a critical disadvantage: the discontinuity of tangential wind and geopotential field when hurricanes develop. This discontinuity will be solved in part II of this series of papers. It was found that this discontinuity is actually not a serious problem and can be removed satisfactorily.

With the fact that these above theories need an initial vortex to start with, this work should be considered as a finding of the analytical solution for the early stage of hurricane development *once the favorable conditions for hurricanes to grow have appeared*. This work does not answer the question "under which conditions does appear such a vortex initially?" It may help us explain why we usually have to create a bogus vortex in numerical models to integrate. This work also provides a method for constructing such a bogus vortex analytically.

# Figure Captions

**Figure 1**. The development of vertical motion with time

**Figure 2a**. The theoretical solution of geopotential with radius: Delta function (1.16)

**Figure 2b**. The observations of wind and geopotential profiles with radius of Hurricane Anita at 500mb surface (Willoughby *et. al.* 1982)

**Figure 3**. The theoretical solution of geopotential with radius: top-hat function (1.19a,b)

**Figure 4**. The theoretical solution of tangential with radius: Delta function

**Figure 5a**. The theoretical profile of homogeneous solution $F_m$ with height

**Figure 5b**. The mean observed structure of tangential wind near the core region of western Pacific typhoon (left panel) and western Atlantic hurricane (right panel) (adapted from McBride 1981).

**Figure 6**. The radius-height profile of the development of theoretical solution of tangential wind with time (time unit is normalized)

**Figure 7**. The development of hurricane Diana at its early stage. (adapted from Willoughby *et. al.* 1990)

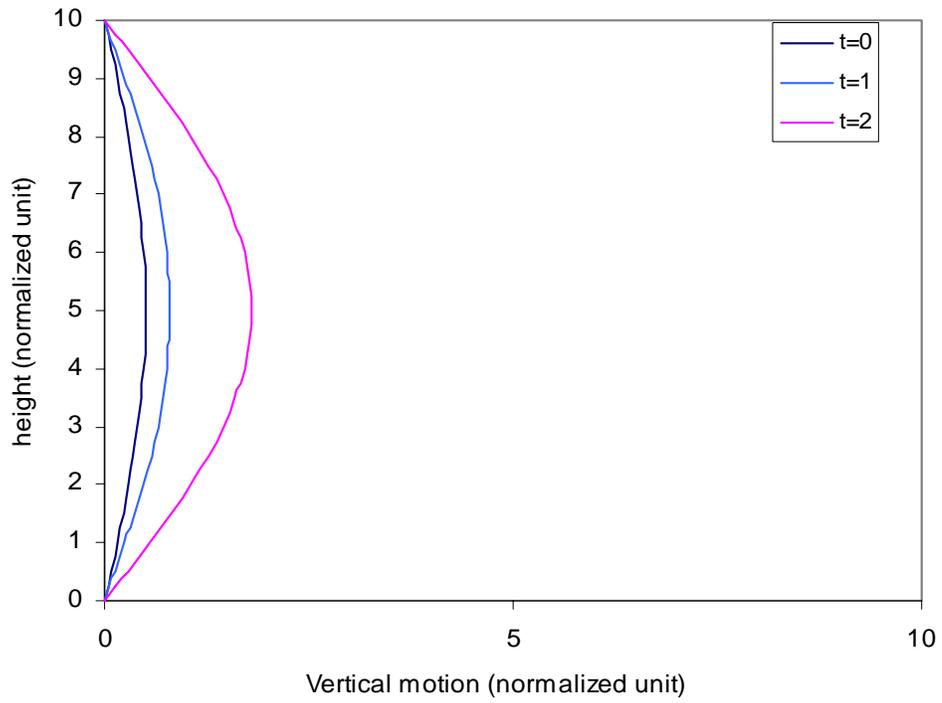

Figure 1. The development of vertical motion with time

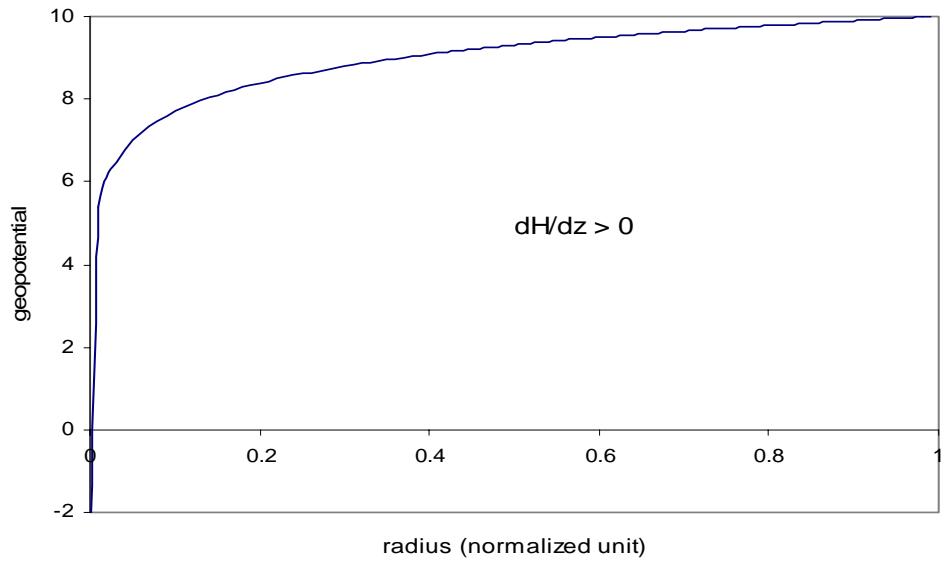

Figure 2a. The theoretical solution of geopotential with radius: Delta function (1.16)

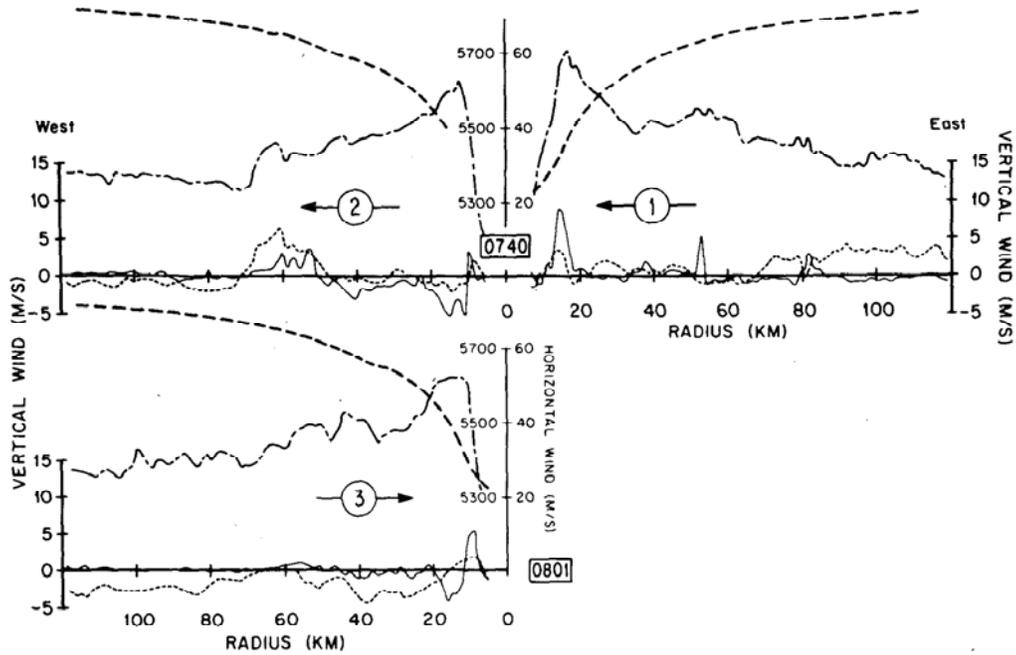

Figure 2.b. The observations of wind and geopotential profiles with radius of Hurricane Anita at 500mb surface (Willoughby *et. al.* 1982)

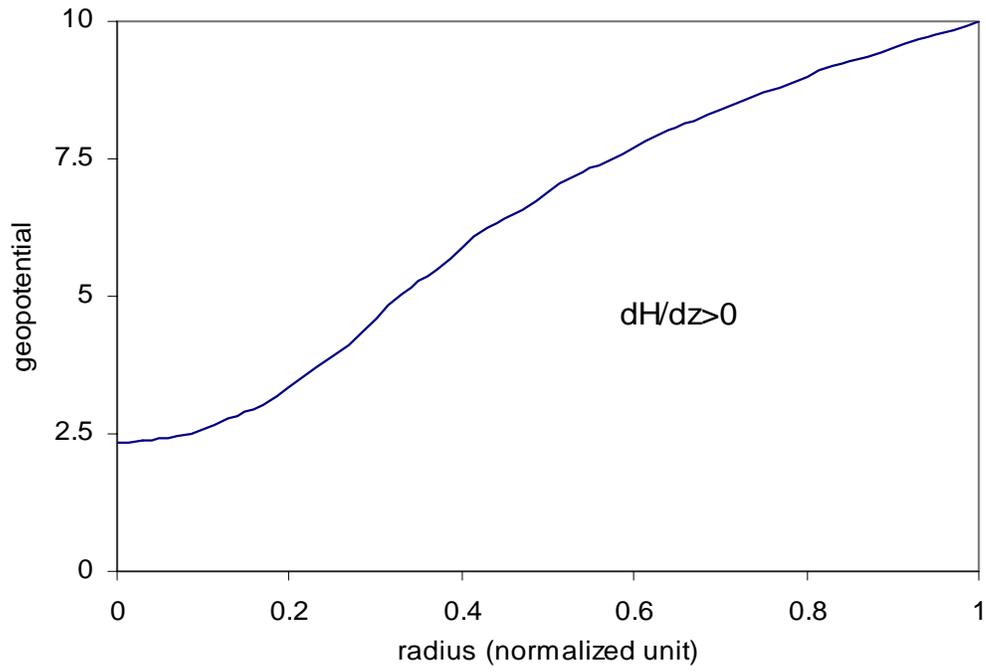

Figure 3. The theoretical solution of geopotential with radius: top-hat function (1.19a,b)

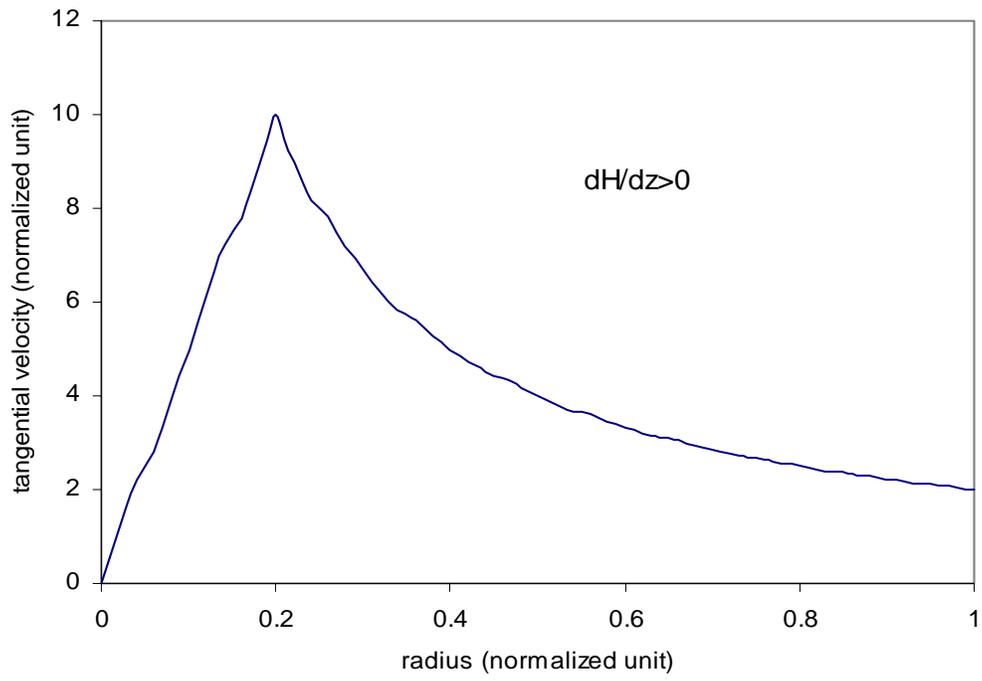

Figure 4. The theoretical solution of tangential with radius: Top-hat function

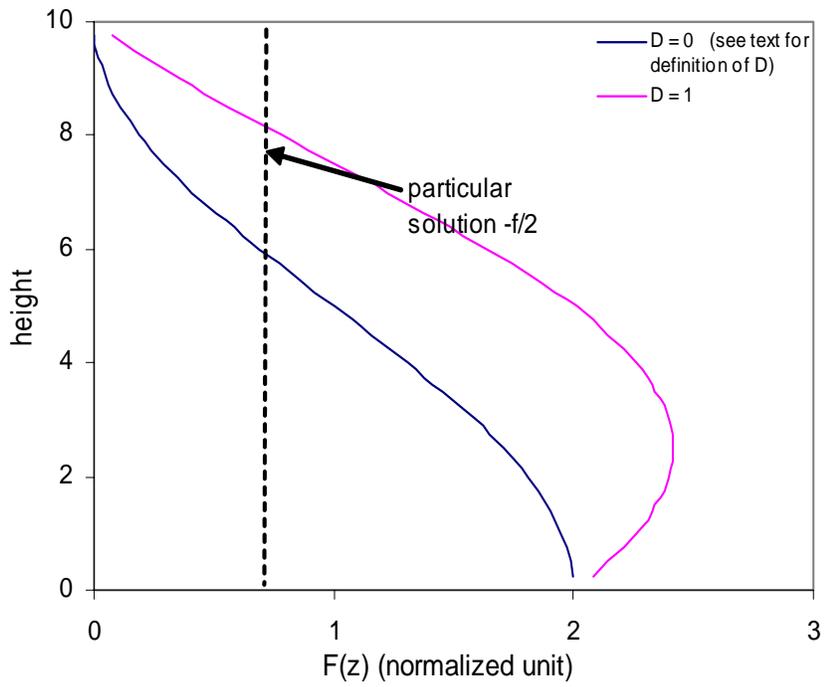

Figure.5a. The theoretical profile of homogeneous solution $F_m$ with height

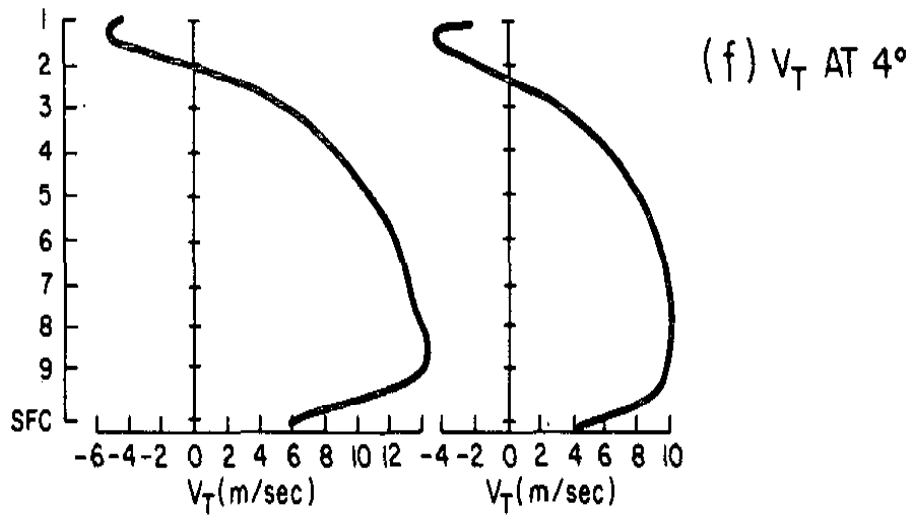

Figure.5b. The mean observed structure of tangential wind near the core region of western

Pacific typhoon (left

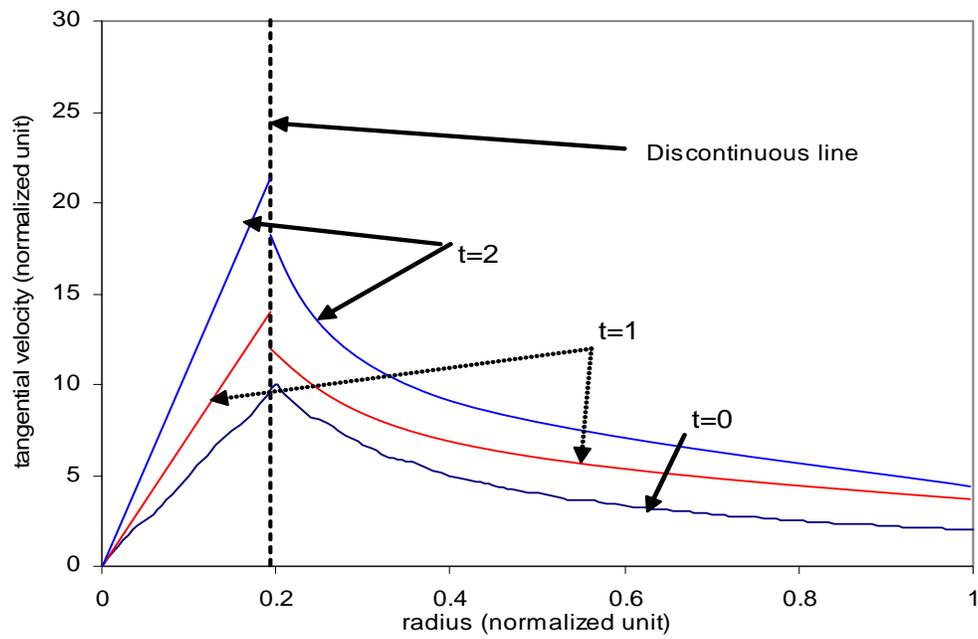

Figure. 6. The radius-height profile of the development of theoretical solution of tangential

wind with time (time unit is normalized)

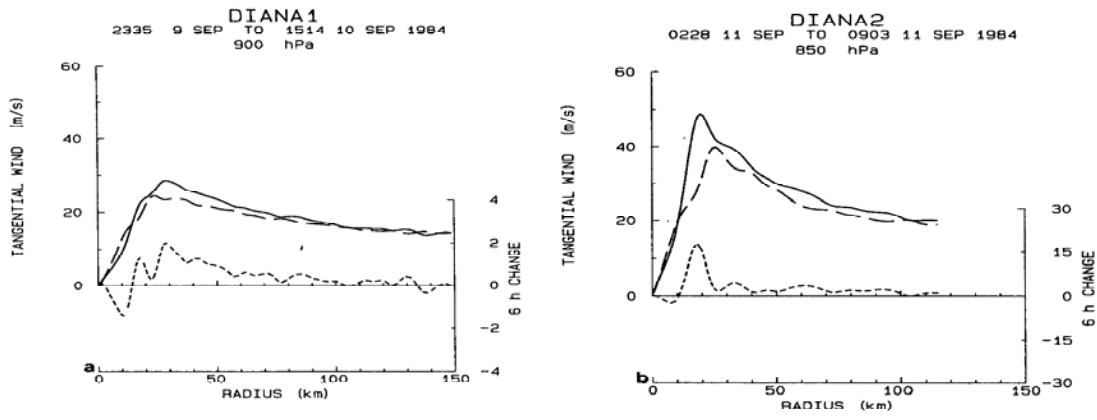

Fig.7. The development of hurricane Diana at its early stage. (adapted from Willoughby *et. al.* 1990)